\documentclass[aps,prb,reprint,groupedaddress]{revtex4-1}

\usepackage{graphics}

\begin{document}


\title{Comparison of dynamical and equilibrium surface spin-reorientations:\\ Inferences on the nature of the phase transitions in the presence of dipole interactions}

\author{G. He}

\author{R. Belanger}
\author{P. H. Nguyen}
\author{D. Venus}
\email{[corresponding author] venus@physics.mcmaster.ca}
\affiliation{Department of Physics and Astronomy, McMaster University, Hamilton, Ontario, Canada}

\date{\today}

\begin{abstract}
The influence of long-range dipole interactions on two dimensional magnetism has been studied extensively in the spin-reorientation transition of ferromagnetic ultrathin films.  Although there is a great deal of experimental information on the perpendicular domain phase that is stabilized by dipole interactions, the transitions to or from the domain phase are subtle and difficult to characterize experimentally.  Magnetic susceptibility measurements show no divergence in the vicinity of the spin-reorientation transition as a function of thickness -- a null result that is difficult to interpret with confidence.  This article reports separate dynamical and equilibrium versions of the reorientation transition in Fe/2ML Ni/W(110) films, using measurements of the magnetic susceptibility as the films are grown.  The dynamical version occurs when increasing the film thickness causes the domain walls to depin, and the system moves from a configuration that minimizes local energetics to one that minimizes global energetics.  The dynamical transition \textit{is} marked by a divergent magnetic susceptibility measured with a field applied along the in-plane W[001] direction.  A comparative study of the two versions of the same spin-reorientation transition aids in the experimental characterization of the effects of dipole interactions on the phase transitions. This comparison confirms the original null result found in magnetic susceptibility measurements of the equilibrium transition; despite its name, the spin-reorientation transition in ferromagnetic ultrathin films has no critical phase transition in either the magnetization or its orientation.
\end{abstract}


\maketitle

\section{Introduction}
Ferromagnetism in two dimensions (2D) is very sensitive to small, perturbing effects.  In fact, because an isotropic 2D system of exchange-coupled spins cannot form a ferromagnetic state at finite temperature\cite{Mermin}, one can say that the existence of ferromagnetism in 2D relies on perturbating effects.  It is therefore not surprising that weak, long-range magnetic dipole interactions have a profound effect on the magnetic states and magnetic phase transitions in 2D.

The study of perpendicularly-magnetized ultrathin films has contributed greatly to understanding the role of dipole interactions in 2D magnetism . This area of research is often termed the ``spin-reorientation transition", because, in simplest terms, the effect of dipole interactions for an infinite planar film can be represented by a constant demagnetization, or shape, anisotropy\cite{Heinrich}.  If a film has perpendicular surface crystalline anisotropy, this is balanced against the shape anisotropy due to short-range dipole interactions to determine the orientation of the magnetization -- perpendicular or in-plane.  As the surface anisotropy varies due to film thickness\cite{Pierce1}, temperature renormalization\cite{Berger1}, or other factors\cite{Dabrowski1}, the balance of  anisotropies may change sign and produce a reorientation of the magnetization between perpendicular and in-plane alignments\cite{Millev1}.  In this picture, most reorientations are second-order phase transitions\cite{[A counter-example of a first-order transition is in: ]Oepen1} and should be marked by a divergence in the magnetic susceptibility.

Long-range dipole interactions complicate this picture substantially by introducing a pattern of magnetic domains in the perpendicularly-magnetized state\cite{Allenspach}.  The periodicity of the pattern is determined by a balance between dipole energy and domain wall energy, and varies exponentially as the anisotropy changes with either temperature or film thickness\cite{Kashuba1,Abanov1}.  Many experimental studies have confirmed the properties of this perpendicular stripe  domain state, including   systematic investigations of  the domain width\cite{Wu1,Meier1}, the domain wall profile and structure\cite{Chen1,Vindigni1}, domain pinning and activation\cite{Venus1,Metaxas1,Kuch1}, and the role of domain pattern defects and fluctuations in its evolution\cite{Libdeh1,Kronseder1,Rodriguez1}.  

However, an understanding of the effect of the domain pattern on the phase transitions themselves remains a difficult and subtle question.\cite{Whitehead,Pighin1}  The presence of perpendicular domains gives a net perpendicular magnetization $M_{\perp}=0$ on a mesoscopic scale, and produces a formally paramagnetic response to a small normal field\cite{Abanov1}.  Whether or not this changes the transitions in a fundamental way in real systems is not obvious.  There is a successful history of studying perpendicularly-magnetized ferromagnetic films using hysteresis loops or ferromagnetic resonance\cite{Bland-Heinrich}, that has established a robust perpendicular magnetization on a microscopic scale -- there may be a delicate question of relative size involved.  Experience with three-dimensional ferromagnets indicates that magnetic domains complicate the analysis of the Curie transition\cite{Arrot} but do not alter the essential character within the critical region.  It is not clear whether or not this will be the case in 2D.

The perpendicular domain pattern also introduces an in-plane orientational order (the pattern) that can be expressed as an order parameter\cite{Whitehead} $O$, given by the relative number of horizontally and vertically displaced nearest-neighbour spins that are aligned. This can further complicate the phase diagram by allowing changes in the pattern symmetry, or ``melting" to a disordered configuration.
 
It is difficult to characterize the nature of the transitions from or to the perpendicular domain phase experimentally.  Magnetic microscopy experiments have been effective in providing evidence of changes in the order of the domain pattern\cite{Vaterlaus1,Portmann1}, but temporal limitations of the imaging technique make it  difficult to approach the critical region and characterize the transitions.  Measurements of the magnetic susceptibility can be made throughout the critical region of the spin-reorientation transition\cite{Arnold3,Venus1,He}, but there is no evidence\cite{He,He2} of a magnetic phase transition from the perpendicular domain state or to the in-plane magnetized state, as would normally be indicated by a divergence in the appropriate component of the magnetic susceptibility\cite{[An exception occurs when the film is coupled to an substrate that is magnetized in-plane : ]Arnold}.  A similar problem is encountered at the thermal transition from the perpendicular domain phase directly to the paramagnetic phase, where domain wall fluctuations make imaging difficult\cite{Won1} and demagnetization effects render magnetic measurements ambiguous or insensitive\cite{Arnold3,Saratz}.

This lack of a marker of a magnetic transition may provide important information, but it is difficult to interpret a null result with confidence.  There are many prosaic reasons that experiments may give a null result, including poor sample preparation, insufficient sensitivity and experimental procedures that are not optimized.  In the present article, we report measurements of the magnetic susceptibility at a dynamical realization of a spin-reorientation transition in Fe/2ML Ni/W(110) ultrathin ferromagnetic films.  The dynamical reorientation occurs in films as they are being grown.  Domain walls are pinned in thinner films and the spin configuration minimizes local energetics.  Domain walls move freely in thicker films and the spin configuration minimizes global energetics.  The dynamical reorientation occurs at the film thickness where the domain walls depin and the spin configuration moves between the locally and globally determined states.  The present experiments show that the magnetic susceptibility \textit{does} diverge at the dynamical reorientation, in the presence of the domain pattern. This permits a comparative analysis of the dynamical and equilibrium versions of the same reorientation transition, and provides insight into the nature of the phase transitions in the presence of dipole interactions.  It turns out that the spin-reorientation magnetic phase transition in ferromagnetic films is not very aptly named, as it does not involve a critical phase transition in either the magnetization or its orientation.
	
%


\section{Magnetic susceptibility near a spin-reorientation transition}
\subsection{Global equilibrium magnetic state}
The following brief summary of the spin-reorientation transition concentrates on simple models that exhibit the symmetries of the system, and define the relevant variables.  The most straightforward model is that of a uniform, planar film of infinite extent, where the order parameters are the uniform magnetization $|M|$, and the angle $\phi$ it makes with the surface normal.  The demagnetization factor $D$ is unity for the magnetization component normal to the film.  For the 4-5 ML films in this study, the bulk, or volume, anisotropy is not expected to play an important role in the reorientation, and is not included.\cite{[Due to this approximation reorientation transitions due to the relaxation of bulk strain are omitted from consideration : ]Farle4}  The following analysis considers varying the thickness of the film at constant temperature $T$.  The notation suppresses the temperature, although it is understood that the magnetic ``constants" renormalize with temperature. Then the Landau expansion of the free energy volume density for the anisotropy can be written as\cite{Fritzsche,Millev1}
\begin{equation}
\label{Landau}
E_{anis}= K_{eff}(\theta) \sin^2\phi + K_4(\theta) \sin^4\phi.
\end{equation}
This uses the convention of $E_{anis}>0$ for a perpendicularly-magnetized film.  The second and fourth order surface anisotropy constants $K_{eff}(\theta)$ and $K_4(\theta)$ depend upon temperature and the average thickness $d=b\theta$, where $b$ is the lattice parameter of the film perpendicular to the substrate, and $\theta$ is the film deposition in monolayers (ML).  In an ultrathin film, the second order effective anisotropy arises from the surface anisotropy energy areal density $K_S$ and the shape anisotropy energy volume density due to short-range dipole interactions, $\Omega=\frac{1}{2}\mu_0DM_{sat}^2$, where $M_{sat}$ is the saturation magnetization.
\begin{equation}
\label{KS}
K_{eff}(\theta)=\frac{K_S}{b\theta}-\Omega.
\end{equation}
A standard minimization of the free energy with respect to $\phi$ shows that if $K_4>0$, a second-order reorientation transition from perpendicular magnetization to a canted state ($0<\phi<\pi/2$) occurs when $K_{eff}$ changes from positive to negative as a function of temperature or coverage.  Eq.(\ref{KS}) gives the deposition when canting begins as\begin{equation}
\label{thetaR}
\theta_R(T)=\frac{K_S}{b\Omega},
\end{equation}
due to the implicit variation of magnetic quantities with temperature.  The effective anisotropy can be rewritten as
\begin{equation}
\label{Ktheta}
K_{eff}(\theta)=\Omega \, \Bigl(\frac{\theta_R(T)}{\theta}-1\Bigr).
\end{equation}

At $\theta_R(T)$, the magnetic susceptibility measured in an in-plane field aligned with the direction in which the magnetization is reorienting, $\chi_{\parallel}=\frac{dM_{\parallel}}{dH_{\parallel}}$, will diverge\cite{Venus1} just as the magnetization begins to cant.  The condition $K_{eff}<-2K_4$ marks the transition from the canted state to uniform in-plane magnetization.  At this point, the magnetic susceptibility measured in a field perpendicular to the film, $\chi_{\perp}=\frac{dM_{\perp}}{dH_{\perp}}$, will diverge.   If $K_4<0$, the phase transition is first-order and the canted state is bypassed.  Then there are no peaks in the magnetic susceptibility. 

If, instead of reorienting, the perpendicular magnetization $|M|\rightarrow 0$ at a Curie transition, then the transition is not observed in $\chi_{\perp}$ because of the demagnetization field.\cite{Venus1}.  

A more complete model includes long-range dipole interactions. Then domains form in the perpendicularly-magnetized state\cite{Kashuba1,Abanov1}.  In equilibrium, the domains form a stripe pattern with a domain density determined by a balance between the energy per unit area $E_W(\theta)$ added when a domain wall is inserted, against the reduction in the long-range dipole energy when a domain is created.  Then the domain density $n^{eq}(\theta)$ is the inverse of the domain width $L$. 
\begin{equation}
\label{neq}
n^{eq}(\theta)=\frac{2}{\pi \ell}\exp[\frac{-E_W(\theta)}{4\Omega b \theta}-1],
\end{equation}
with $E_W(\theta)=4\sqrt{\Gamma K_{eff}(\theta)}$, and the domain wall width $\ell(\theta)=\pi \sqrt{\Gamma/K_{eff}(\theta)}$.  $\Gamma$ is the exchange stiffness.  This expression is valid so long as the domain wall width is significantly smaller than the  domain width.  Using eq.(\ref{Ktheta}), this can be expressed as
\begin{equation}
\label{ntheta}
n^{eq}(\theta)=\frac{2}{\pi \ell}\exp[-\sqrt{\frac{\Gamma}{\Omega b^2}}\frac{\sqrt{\theta_R(T)-\theta}}{\theta^{3/2}}-1].
\end{equation}
The magnetic susceptibility of the perpendicularly-magnetized domain state is a due to co-ordinated motion of the domain walls, where the domains parallel to a field applied normal to the film grow, and those that are antiparallel shrink.  The equilibrium magnetic susceptibility in a small perpendicular field is proportional to the domain width\cite{Abanov1},
\begin{equation}
\label{chi}
\chi_\perp^{eq}(\theta)=\frac{2}{\pi^2 b\theta} \, \frac{1}{n^{eq}(\theta)},
\end{equation}
and falls exponentially with increasing deposition or temperature.  The exponentially decreasing region of susceptibility measurements can be analyzed by approximating the pre-exponential in eq.(\ref{chi}) as a constant $\chi_0$, and plotting
\begin{equation}
\label{chiplot}
\theta^3 \, [\ln{\chi_\perp^{eq}}-\ln{\chi_0}-1]^2=\frac{\Gamma}{\Omega b^2}\, (\theta_R(T)-\theta).
\end{equation}
Then $\theta_R(T)$ can be determined by linear extrapolation from a region where corrections due to the saturation of the domain wall width\cite{Abanov1}, higher order anisotropy\cite{Stickler} $K_4$, and the Dzyaloshinskii-Moriya interaction\cite{Meier1} are negligible.

\subsection{Local metastable magnetic state}
\begin{figure}
\scalebox{.55}{\includegraphics{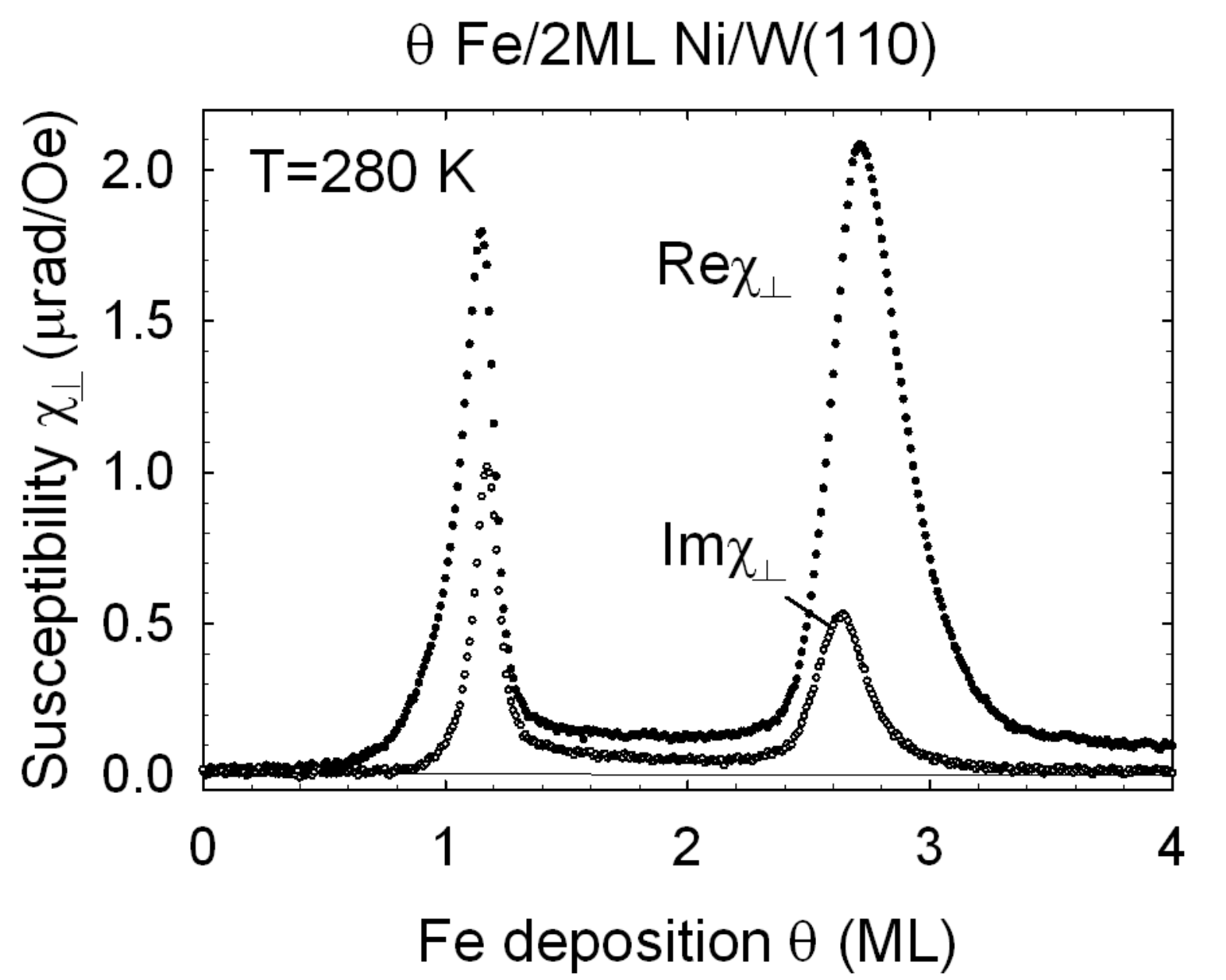} }
\caption{\label{fig1}The magnetic susceptibility of an Fe/2ML Ni/W(110) film measured in real time as the Fe was deposited at 280 K. The oscillatory field of 2 Oe was applied normal to the surface.  Re$\chi_\perp$ is the response in phase with the applied field, and Im$\chi_\perp$ is the dissipative response out of phase.  The experimental methods are presented later in the article.  The two prominent peaks in the susceptibility are due to the response of the metastable, local reorientation transition (lower deposition) and the response of the equilibrium domain phase in the global reorientation transition (higher deposition).}
\end{figure}

Fig.\ref{fig1} show the magnetic susceptibility, measured with a small field normal to the surface, while an Fe film is being grown at 280 K on a 2ML Ni/W(110) substrate.  Similar measurements are analyzed quantitatively in ref.(\onlinecite{He}).  The two peaks are the magnetic response at the local  (near 1 ML) and the global  (near 3 ML) realizations of the same reorientation transition.

The peak at higher deposition is due to the response of the equilibrium domain state outlined in the previous section, with the exponential decrease (above about $\theta$=2.8 ML in this example) due to the change in equilibrium domain density as given by eq.(\ref{ntheta}) and (\ref{chi}).  On the left hand side of this peak, the domain walls become progressively pinned by structural defects, and respond with a relaxation time $\tau$ given by
\begin{equation}
\label{tau}
\tau=\tau_0 \exp(\frac{E_a}{kT}),
\end{equation}
where $E_a$ is an activation energy and $\tau_0$ is a characteristic time between ``attempts" to escape the pinning site.

In a model by Bruno \textit{et al.}\cite{Bruno1}, the activation energy $E_a$ is due to pinning at the steps at monolayer changes in thickness.  They find that the mean of the distribution of activation energies is given by
\begin{equation}
\label{Eadef}
E_a(\theta)=\frac{\zeta b \theta}{E_W(\theta)} \Bigl(\frac{\partial}{\partial \theta} E_W(\theta) \Delta \theta\Bigr)^2,
\end{equation}
where $\zeta$ is the mean spacing of pinning sites.  Using eq.(\ref{Ktheta}) and $\Delta\theta$= 1 ML, this can be rewritten as
\begin{equation}
\label{Eatheta}
E_a(\theta)=\frac{\zeta b\,  \sqrt{\Gamma \Omega}\,\theta_R^2(T)}{\theta^{3/2}(\theta_R(T)-\theta)^{3/2}}.
\end{equation}
Note that the pinning energy increases as the deposition, or average thickness, decreases. When the relaxation time for pinning increases at lower deposition, it reduces the measured susceptibility, so that a peak is formed near 2.7 ML.  It can be seen in fig.(\ref{fig1}) that the dissipation, as represented by Im$\chi_\perp$, is largest on the left hand side of the peak, where the domain walls move between the pinning sites.  Eq.(\ref{chi}), (\ref{tau}) and (\ref{Eatheta}) provide an excellent quantitative description of the peak in the measured susceptibility at higher deposition\cite{He}.

The peak at lower deposition in fig.(\ref{fig1}) is due to a metastable realization of the reorientation transition.  For $\theta<$ 2.2 ML in this example, the domain walls are pinned.  This means that, as the initial layers of the film are deposited, the domain configuration cannot respond to the global average anisotropy $K_{eff}$, which is determined by the average thickness $d=b\theta$.   Since, in this system, a 3 ML Fe film has in-plane effective anisotropy and a 2 ML Fe film has perpendicular effective anisotropy, each 3 ML island reorients locally and independently\cite{Weber1,Kubetzka1,Gabaly1}.  On the left hand side of the peak, the 3rd layer islands are small and each is ringed by a partial domain wall.  The spins in the partial wall are ``soft" to a perpendicular field because the in-plane anisotropy energy opposes the exchange coupling to the perpendicular spins outside the island.  The susceptibility increases as the islands grow.   Once the islands have a radius greater than the size of a $90^o$ domain wall, the in-plane anisotropy and the exchange coupling with the in-plane spins in the interior of the island are mutually reinforcing.  This stiffens the response to a perpendicular field, and the susceptibility peak is cut off.  For this reason, the dissipative response in Im$\chi_\perp$ is on the right hand side of this peak in Re$\chi_\perp$.    This model is developed in ref.(\onlinecite{He}) and shown to give an excellent quantitative description of the first peak in the experimentally measured susceptibility.

\subsection{Dynamical reorientation}
The existence of a local and global reorientation implies that there must be third transition at an intermediate deposition.  
At low deposition when domain walls are pinned, the spins in the 3rd layer Fe islands move from perpendicular to in-plane alignment to minimize the local energy.  At high deposition, the equilibrium transition requires the free movement of domain walls to access a global minimum in the free energy which produces an ordered perpendicular domain state.  At the intermediate coverage where the domain walls depin, there will be a dynamical reorientation transition where the system moves from the local to global energy minimum.  In this transition, the spins in the 3rd layer Fe islands must revert to perpendicular alignment.   Since  depinning is a dynamical response, the measured susceptibility depends upon the time scale of the measurement.  The measurements in fig.(\ref{fig1}) were made using a small field oscillating at 210 Hz.  The films were grown at a very slow rate of about $2\times10^{-3}$ ML/s.  Thus, the domain configuration can adapt to depinning in the slowly changing film structure, even when it cannot respond to the oscillating field.  

A simple relaxation model can be used to estimate the intermediate depinning deposition $\theta_d(T)$ at which the
 dynamical reorientation is expected to occur.  Consider the situation where the equilibrium domain density must change in response to a change in the film thickness.  The instantaneous domain density $n(\theta)$ relaxes to the equilibrium domain density $n^{eq}(\theta)$ according to\cite{Libdeh2}
\begin{equation}
\label{relax}
\frac{dn(\theta)}{dt}=\frac{-1}{\alpha \tau} [n(\theta)-n^{eq}(\theta)].
\end{equation}
$\tau$ is the same relaxation time as in eq.(\ref{tau}), since the pinning sites are the same.  However $\alpha$ is a numerical factor that takes into account the differences in geometry and scale in the response of a mesoscopic domain pattern and the response of a small section of domain wall.  A previous experimental study\cite{Libdeh2} found that $\alpha\approx10^{5.5}$ for the relaxation of the domain density in response to a change in temperature.

For an estimate of the depinning deposition, the growth rate $R=\frac{d\theta}{dt}$ is used to convert the time rate of change to the coverage rate of change.  If the system is not too far from equilibrium, the functional forms of the instantaneous and equilibrium density will be similar, and a first approximation is to replace $\frac{dn}{d\theta}$ in eq.(\ref{relax}) by $\frac{dn^{eq}}{d\theta}$.  Using eq.(\ref{ntheta}) then yields
\begin{equation}
\label{depin}
n(\theta)=n^{eq}(\theta)[1-R\alpha \tau g(\theta)],
\end{equation}
where $g(\theta)$ is the derivative, with respect to $\theta$, of the argument of the exponential in $n^{eq}(\theta)$ in eq.(\ref{ntheta}).  Since $n(\theta)$ and  $n^{eq}(\theta)$ cannot be negative, it is not possible for the domain density to relax toward the equilibrium configuration once the expression in square brackets passes through zero.  Therefore, the domain relaxation is pinned when
\begin{equation}
\label{thetadepin}
R\alpha\tau g(\theta_d)=1.
\end{equation}
Using eq.(\ref{tau}) and (\ref{Eatheta}), this can be expressed as
\begin{equation}
\label{thetad}
\frac{T_0 \, \theta_R^2(T)}{T\theta_d^{3/2}(T)(\theta_R(T)-\theta_d(T))^{3/2}}=1,
\end{equation}
where 
\begin{equation}
\label{T0}
T_0=\frac{\zeta b \, \sqrt{\Gamma \Omega}}{k\ln[R\alpha \tau_0 g(\theta_d)]^{-1}}.
\end{equation}
Eq.(\ref{thetad}) can be rearranged to a quadratic form for an estimate of the depinning deposition where the dynamical reorientation is expected to occur:
\begin{equation}
\label{quad}
\frac{\theta_d(T)}{\theta_R(T)} -[\frac{\theta_d(T)}{\theta_R(T)}]^2=[\frac{T_0}{T\theta_R(T)}]^{2/3}.
\end{equation}

\section{Experimental methods}
The experiments were performed in the same manner as those described in ref.(\onlinecite{He}).  The following short summary is abstracted from that publication, with emphasis on any changes in procedure.

Measurements of the magnetic susceptibility were made \textit{in situ} as an ultrathin film was grown on a W(110) single crystal substrate in ultrahigh vacuum.  The sample holder\cite{Venus2} was equipped with electron beam heating for flashing to high temperature, radiative heating for temperature control, and a liquid nitrogen reservoir for cooling.  The sample could be rotated through polar and azimuthal angles, so that any in-plane crystalline axis could be aligned with an in-plane pair of magnetic field coils, and with the scattering plane of the laser beam used for the magneto-optic measurements.  A second coil attached to the holder generated a field normal to the sample surface for measurements of $\chi_\perp$.  The substrate cleanliness was confirmed using low energy electron diffraction and Auger electron spectroscopy (AES).

The films were formed by evaporation from a pure wire.  Electrons thermally emitted from a hot filament inside the evaporator\cite{Jones1} were accelerated by 1.75 kV and bombarded the tip of the wire.  The evaporated atoms were collimated by two apertures and formed a beam directed at the substrate crystal.  The evaporator was supported in an adjustable tripod, so that the direction of the atomic beam could be finely adjusted and made to coincide with the region of the film probed by the laser used for magneto-optic Kerr effect (MOKE) measurements.  AES was used to iteratively adjust the evaporator direction to ensure a uniform film over a region about 9 mm$^2$ on the substrate.

The second collimating aperture in the evaporator was electrically isolated.  Because a certain fraction of the evaporate atoms striking it are ionized,  an ion current of order nA could be measured using an electrometer.  Fine adjustments of the wire position were used to keep the monitor current constant and thus ensure a constant deposition rate.  The deposition rate was calibrated by a sequence of accumulating depositions, where the film was annealed to 600K and an W Auger spectrum was measured after each step in deposition.  For Fe/W(110) and Ni/W(110), a plot of the W Auger attenuation vs. deposition time shows a clear break in slope at 1 ML that was used to calibrate the monitor current\cite{[Examples can be seen in ]Fritsch1,*Jones2}.  The stability of the evaporator calibration and  deposition rate over the 20 to 30 minutes required to measure a susceptibility curve during growth was checked\cite{He} by growing Fe films directly on W(110) at 450 K, the known Curie temperature of 2ML Fe films magnetized in-plane, while measuring $\chi_{1\overline{1}0}$.  The peak in the susceptibility at the transition was then used to calculate the average deposition rate.  These tests illustrated that thickness calibrations are accurate to $\pm5$\% over the range of growth rates used.

The magnetic susceptibility of the film was determined with a MOKE apparatus\cite{Arnold2,Arnold4} using a linearly polarized HeNe laser.  Details of the optical arrangement, alignment procedures, sensitivity and conversion of the raw data to magnetic susceptibility can be found in ref. (\onlinecite{Arnold2}) and (\onlinecite{Arnold4}).   The laser beam entered through a UHV window, scattered at 45$^o$ from the substrate normal, and exited through a second UHV window.  Compensation techniques were used to retain linear polarization after the magneto-optical Kerr rotation.  The beam then passed through a polarizing crystal to isolate the rotated component of the light, and was detected by a photodiode.  An a.c. field of 2.0 Oe and 210 Hz was generated by either the in-plane or normal coils, depending upon the experiment, and lock-in detection was used to isolate the signal at the frequency of the field.  The susceptibility is measured directly in units of $\mu$rad/Oe, with the real and imaginary parts obtained simultaneously as the in-phase and out-of-phase components from the lock-in amplifier.

Measurements of the reorientation transition were made for Fe deposition on a substrate of 2ML Ni/W(110).  The nickel film was annealed to 600 K after the deposition of 1 ML to cause wetting of the substrate.  In this system\cite{Johnston1}, the Ni layers create a slightly strained f.c.c. (111) surface template with atomic spacing very close to that of bulk Ni, and an in-plane magnetization.  Subsequent pseudomorphic Fe deposition creates a system with perpendicular anisotropy.  Thicker Fe films reorient to an in-plane magnetization along the [001] in-plane direction of the underlying W(110) crystal.   The susceptibility was measured using an a.c. field directed along the normal, or the appropriate in-plane direction for measurements of $\chi_\perp$, $\chi_{001}$, or $\chi_{1\overline{1}0}$.

It is important to remember that each susceptibility measurement in this study represents the growth of a new film.  Because the film growth is reproducible to a great degree, comparisons of susceptibilities using different field geometries are made for different films grown on the same, or successive, days.  However, even though two data traces are often shown on the same plot, the curves cannot be expected to align to greater precision than the accuracy of the thickness calibration.
\begin{figure}
\scalebox{.64}{\includegraphics{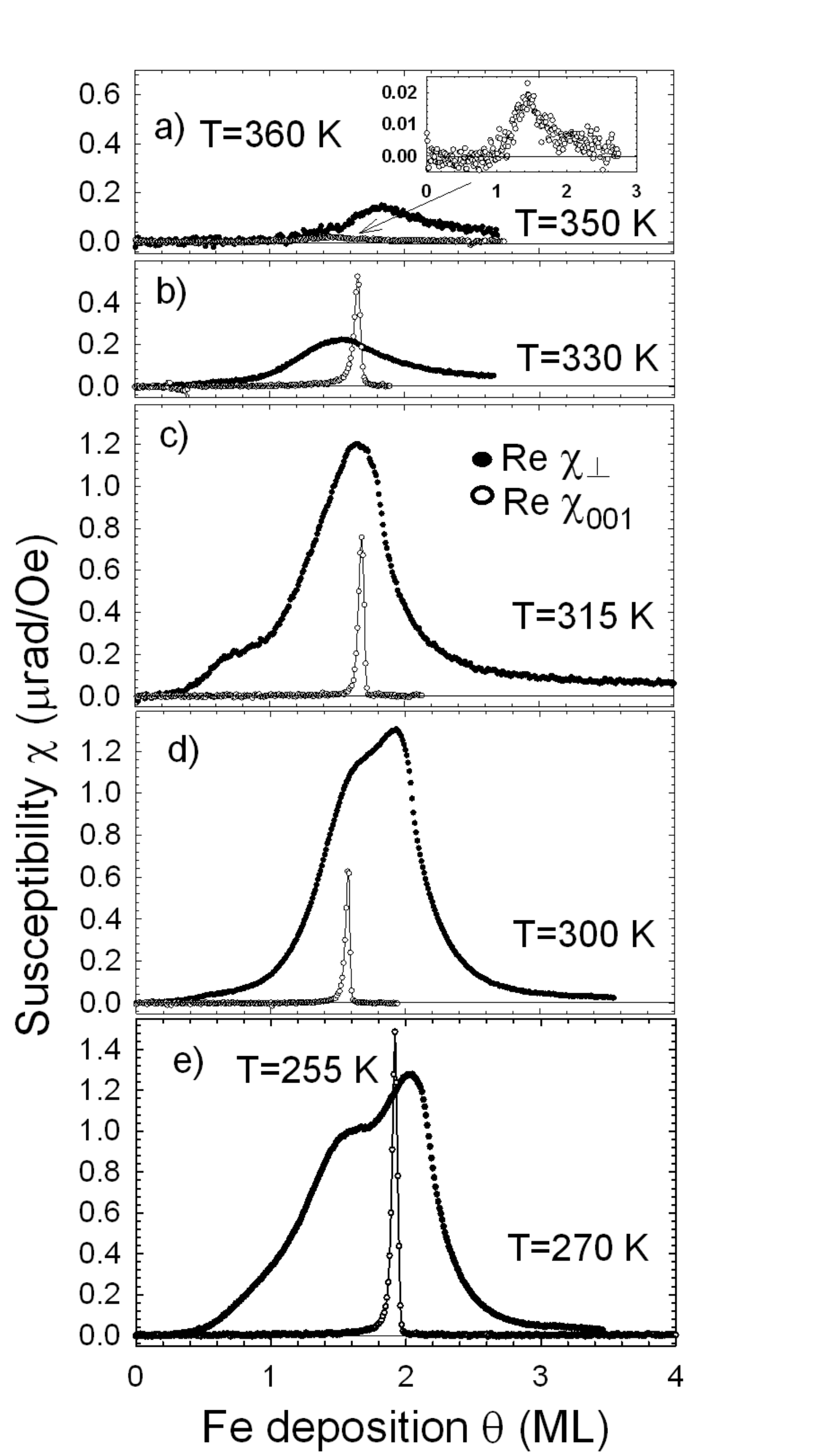}}
\caption{\label{fig2}The magnetic susceptibility measured as the Fe layer is grown on a 2ML Ni/W(110) substrate.  In each panel, measurements of $\chi_{\perp}(\theta)$ made with a field normal to the surface, and of $\chi_{001}(\theta)$ made with a field along the in-plane direction W(001), are shown using the solid and open symbols, respectively.  The temperature at which the measurements were made is indicated on the right of each panel.  In cases where $\chi_{001}(\theta)$ was measured at a slightly different temperature, this is indicated by the temperature on the left.  The inset to part a) shows $\chi_{001}(\theta)$ on an expanded scale.}
\end{figure}

\section{Results and analysis}
The Fe layer was grown at constant temperature on a series  of films while the magnetic susceptibility was measured.  Fig. (\ref{fig2}) shows a collection of these measurements of Re$\chi_\perp$ (solid symbols) and Re$\chi_{001}$ (open symbols).  Measurements of $\chi_{1\overline{1}0}$ produced no signal above the level of noise and are not shown.

The temperature noted on the right hand side of each panel gives the temperature at which $\chi_\perp$ was measured.  If there is a temperature noted on the left hand side of the panel, then it refers to a nearby temperature at which $\chi_{001}$ was measured.  Recall that in all cases the two curves in the same panel where measured during the growth of different films.

The measurements of $\chi_\perp$  are consistent with the previous study of many such films during growth\cite{He}.  When the growth temperature during measurement is below 305 K, the curve is most likely to exhibit a single strong peak with a prominent shoulder on the low-deposition side.  When the growth temperature is above 325 K, the curve is most likely to have a single peak.  In the intermediate range of growth temperature, the curve is most likely to exhibit two well-separated peaks, as in fig.(\ref{fig1}), but a prominent peak with a well-separated shoulder at lower deposition, as in fig.(\ref{fig2}c), occurs in about one quarter of the measurements.  These systematic changes with temperature are attributed to differences in the growth dynamics of the Fe films as a function of temperature;  specifically, to observe two well-separated peaks requires that the Fe adatom mobility is large enough to permit aggregation on existing nucleated 3rd layer islands, but small enough to prohibit hopping into vacancies in the 2nd Fe layer\cite{He}.  The lack of two well-separated peaks at other temperatures does not mean that the local reorientation has not taken place;  rather the island distribution is not always optimal for observing the local reorientation with $\chi_\perp$.  All measurements of $\chi_\perp$ are consistent with the combined response of the local and global realizations of the spin-reorientation transition.

The measurements of $\chi_{001}$ represent, to our knowledge, the first observation of a narrow, divergent susceptibility peak of a perpendicularly-magnetized system within the stripe domain phase.  The peak is prominent up to at least 330 K, with a normalized full width at half maximum, $\Delta\theta_{001}/\theta_{001}\approx$0.03.  This width is consistent with measurements of the diverging susceptibility in previous measurements of second-order  Curie transitions\cite{Dunlavy1} and percolation transitions\cite{Dixon} in ultrathin films.  At 350-360 K, the peak broadens considerably and is greatly diminished in amplitude.  Measurements taken at 380 K show no signal above the noise in either $\chi_\perp$ or $\chi_{001}$.

A divergent peak in $\chi_{001}$ can, in principle, represent a number of different magnetic phase transitions.  It could indicate a percolation transition, where the isolated islands magnetized in-plane form a connected, coherent in-plane magnetic network once the deposition passes a certain threshold.  However, since the peaks occur at about $\theta$=1.5 ML, it is not possible that islands in the 3rd layer have percolated.  Alternatively, the peak  could indicate a Curie transition between in-plane ferromagnetism along the [001] direction and a paramagnetic state.  This does not make sense in the present context, since it implies a paramagnetic state across a wide range of coverages where $\chi_\perp$ clearly indicates that perpendicular magnetic domains persist. 

The only self-consistent explanation is a reorientation transition.  There is strong qualitative evidence supporting this conclusion.   First, a local reorientation of the moments on 3 layer Fe islands at lower deposition, as in fig.(\ref{fig1}), has created a population of in-plane moments that can reorient.  Second, the peak is measured in $\chi_{001}$, which is the expected axis of reorientation, because it is the in-plane easy axis for ferromagnetism in this system\cite{Johnston1}.  Third, the asymmetric shape of the peak indicates that, for an applied field along the [001] direction, the initial state at lower deposition is in-plane and the final state at higher deposition is perpendicular.  This scenario is then consistent with the subsequent evolution of the perpendicular equilibrium domain state seen in $\chi_\perp$ at higher deposition.
\begin{figure}
\scalebox{.5}{\includegraphics{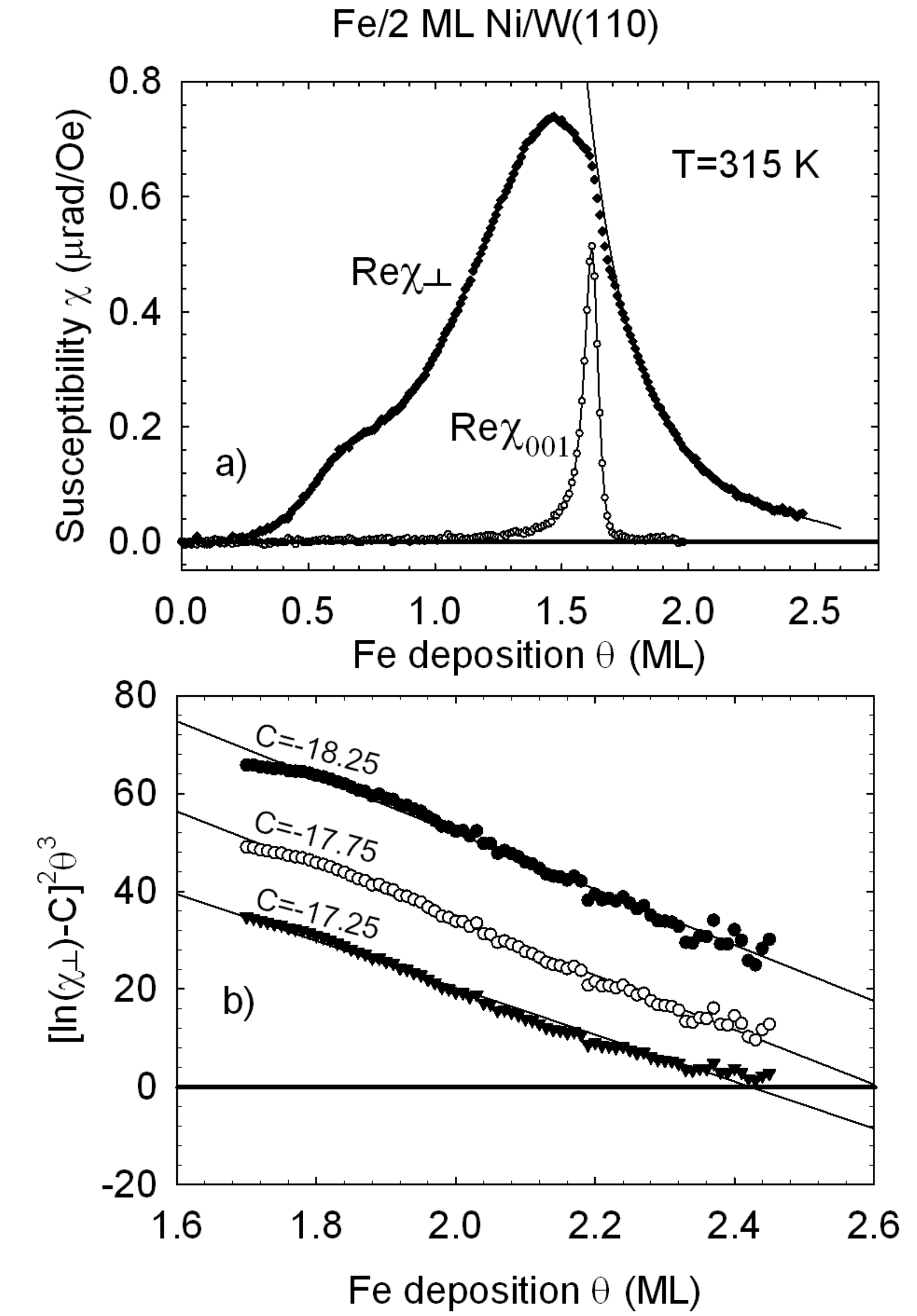} }
\caption{\label{fig3}a) Measurements of the perpendicular (solid symbols) and in-plane (open symbols) susceptibility, made as films were grown at 315 K, are shown as a function of the Fe deposition. b) The high deposition side of $\chi_{\perp}(\theta)$ is analysed according to eq.(\ref{chiplot}), using different values of the constant $C$.  The middle curve represents the best fit, and the extrapolation to cross the deposition axis determines $\theta_R$.  The fitted curve is shown in part a) as the solid line through the solid symbols.}
\end{figure}

To test whether or not $\chi_{001}$ marks a dynamical reorientation tied to the depinning of the perpendicular domain structure, the data are analysed according to the quantitative model developed in Section II.  The first step is to determine $\theta_R(T)$ experimentally.  Fig.(\ref{fig3}a) presents the analysis of a second pair of susceptibility measurements made for Fe films grown at 315 K.  In part b) of the figure, $\chi_\perp$ between 1.70 and 2.45 ML Fe is plotted according to eq.(\ref{chiplot}), for different choices of the parameter $C=1 +\ln\chi_0$.  ($C$ depends on the units of the susceptibility.)  The excellent linear fits confirm that the susceptibility is the response of the perpendicular domain state.  The best linear least-squares fit is obtained for the value $C=-17.75$, with the least-squares residuals rising by 25\% for the neighbouring values of $C$ included in the plot.  The intercept of the best fit line with the deposition axis gives $\theta_R$(315 K)=2.61 ML for this data set.  The fitted susceptibility is given by the solid line through the solid points in fig.(\ref{fig3}a), and represents the data very well.  The peak of $\chi_{001}$ (open symbols) occurs at $\theta_{001}$(315 K)=1.62 ML.

\begin{figure}
\scalebox{.7}{\includegraphics{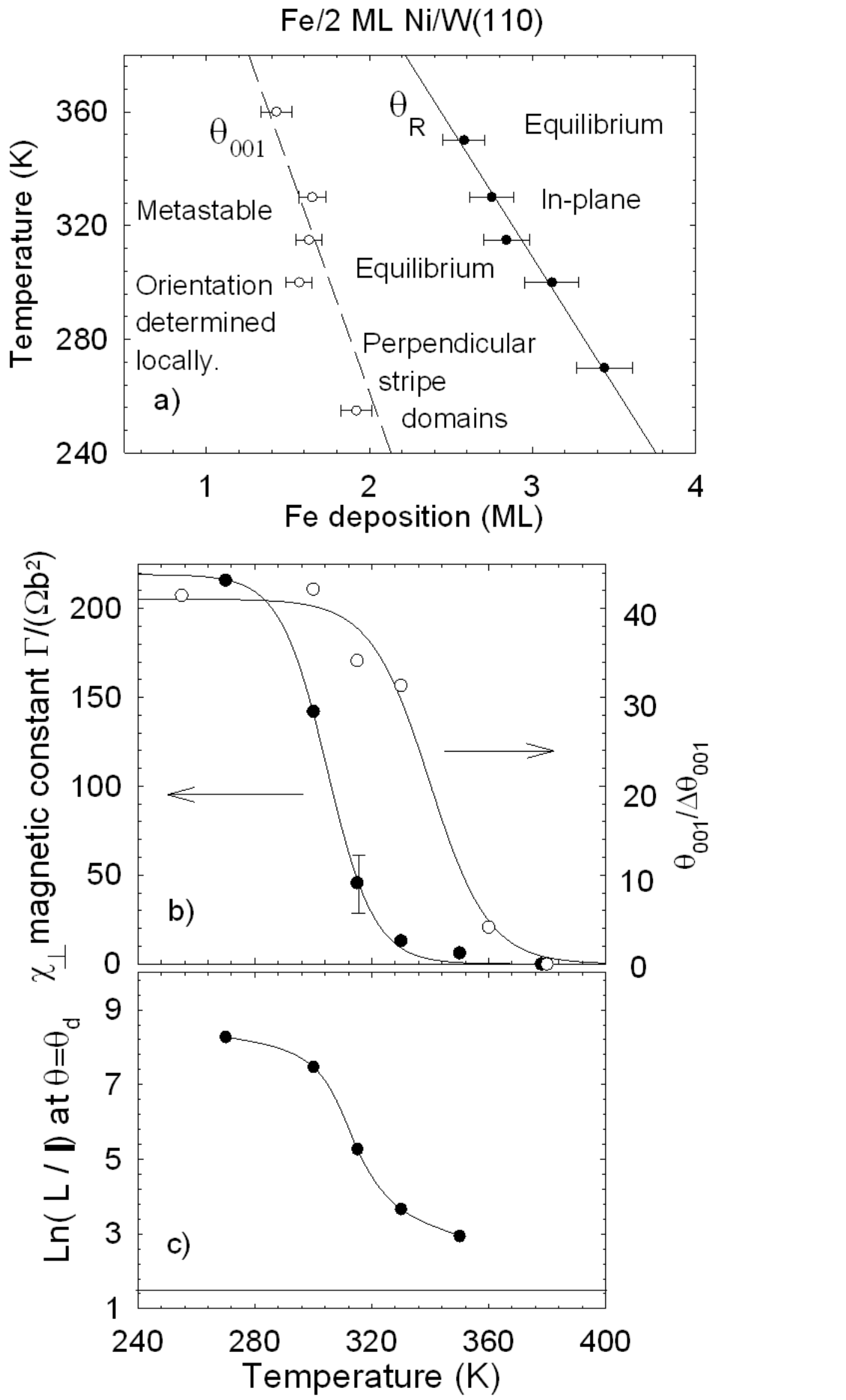} }
\caption{\label{fig4}a) The depositions $\theta_R$ (solid symbols) and $\theta_{001}$ (open symbols) for the data in fig.(\ref{fig2}), as determined through an analysis as in fig.(\ref{fig3}), are plotted in the temperature -- deposition plane. The data at 315 K is the average of five independent measurements; the others are single measurements.  Error bars are due to the thickness calibration.  The solid line is the linear fit given in eq.(\ref{thetaRfit}).  The dashed line is a fit to eq.(\ref{quad}) using the single parameter $T_0=114\pm6$K.  Regions of the phase diagram are labelled in accord with a dynamical reorientation transition at $\theta_{001}$ and an equilibrium reorientation transition at $\theta_R$.  b) The fitted, temperature dependent magnetic constants derived from Re$\chi_\perp$, representing the slope of the line as in fig.(\ref{fig3}b), are plotted as solid symbols (left hand scale).  The error bar for 315 K is the standard deviation for five independently measured films.  The open symbols (right hand scale) are the inverse of the peak width, $\theta_{001}/\Delta \theta_{001}$, of the divergent peaks in $\chi_{001}$ in fig.(\ref{fig2}).  The fitted sigmoidal functions are a guide to the eye. c) The data in parts a) and b) are used in eq.(\ref{ntheta}) to calculate the ratio of the domain width to the domain wall width $L/\ell$ at $\theta_{001}$.  This is shown on a logarithmic scale vs. temperature, where the horizontal line shows the minimum value.}
\end{figure}

Similar analysis of five such pairs of measurements of $\chi_\perp$ and $\chi_{001}$ for films grown at 315 K give average values of $\theta_R$(315 K)=2.84$\pm0.18$ ML and $\theta_{001}$(315 K)=1.63$\pm0.11$ ML (uncertainties are standard deviations).  These uncertainties are indistinguishable from the $\pm5\%$ uncertainty in the thickness calibration.  Therefore, all five measurements are internally consistent.   Fig.(\ref{fig4}a) shows these values on a plot of measurement temperature vs. Fe deposition as solid and open symbols, respectively.  Data points derived from similar fits to the rest of the measurements in fig.(\ref{fig2}) are also included.  The points at these additional temperatures are single measurements with the uncertainty given by the thickness calibration. The solid points in fig.(\ref{fig4}b) are the fitting constant $\Gamma/(\Omega b^2)$, which is the slope of fits such as those in fig.(\ref{fig3}b), at each temperature.  The open points are the inverse of the width $\chi_{001}$ for the data in fig.(\ref{fig2}).  Fig.(\ref{fig4}c) uses the data in parts a) and b) of the figure to calculate the ratio $L/\ell$ at $\theta_{001}$, using eq.(\ref{ntheta}). These will be discussed in the next section.

The solid line in fig.(\ref{fig4}a) is the least-squares linear fit
\begin{equation}
\label{thetaRfit}
\theta_R(T)=(6.4\pm0.3) - (1.1\pm0.1)\times10^{-2}\,T,
\end{equation}
where $\theta_R$ is in ML and temperature is in K.  This implies a linear renormalization of the surface anisotropy with a constant temperature coefficient $\lambda$, as has been found in previous studies\cite{Won1}.  The fitted values of $K_S(T=0)/b\Omega$=6.4 and $\lambda/b\Omega$=0.011 are consistent within uncertainty with a previous study of the domain phase in this system as a function of temperature\cite{Libdeh3}.  Despite all these indications of a reorientation transition from the equilibrium domain phase at $\theta_R(T)$, no divergent susceptibility associated with this transition is observed\cite{He,He2} in $\chi_\perp$ or $\chi_{001}$.

With the experimental expression for $\theta_R(T)$ established, it is possible to test if the peak observed in $\chi_{001}$ is correlated to the depinning of the domain walls in the film, as described by eq.(\ref{quad}). The dashed line in fig.(\ref{fig4}a) is the result of a single parameter least-squares fit to the peak positions that yields the parameter $T_0=114\pm6$K.  This value can be compared to that predicted by  eq.(\ref{T0}).  Using $\zeta$=50 nm\cite{He},  $R=2\times10^{-3}$ ML/s, $\tau_0=10^{-9}$ s, $\alpha=10^{5.5}$, $g(\theta)\approx3$ and the bulk Fe values\cite{Chikazumi} of $\Gamma$ and $\Omega$, results in $T_0\approx 300$.  This is significantly higher than the fitted value, but the discrepancy is consistent with the temperature renormalization of the magnetic constants from their bulk values, as in fig(\ref{fig4}b).

In summary, a single parameter fit gives an excellent representation of the position of the peak in $\chi_{001}$.  This is quantitative evidence that the peak occurs when the domain walls depin.  Taken together with the qualitative evidence, this is a strong case that there is a dynamical reorientation from a mixed metastable state of locally determined in-plane or perpendicular magnetic alignment to one of perpendicular alignment, and that this occurs when the domains relax to a configuration determined by global energetics.

\section{Discussion}
The measurements of dynamical and equilibrium versions of the same reorientation transition provides an opportunity for a comparative analysis and an experimental characterization of the individual phase transitions.

First, the observation of a divergence in $\chi_{001}$ at the dynamical transition indicates that it is a second-order transition.  According to eq.(\ref{Landau}), this means that  $K_4>0$ at $\theta_{001}(T)$.  Since it is highly unlikely that the sign of $K_4$ is different at the nearby deposition $\theta_R(T)$, the transitions at the equilibrium reorientation will also be second-order.  This proves that even though the measured magnetic susceptibilities do not diverge at the equilibrium transition, is not because it is a first-order transition.
 
Second, the absence of a response to an in-plane field along W(1$\overline{1}$0) at the dynamical transition confirms a strong in-plane anisotropy along W(001) that will be essentially unchanged at the equilibrium transition.  This means that the type of domain pattern melting, or transition from a stripe pattern to a tetragonal pattern, that has been observed in systems with four-fold\cite{Vaterlaus1,Portmann1} or polycrystalline\cite{Stickler,Fromter} in-plane symmetry will be strongly discouraged.  The present films are more analogous to those studied by Bergeard \textit{et al.}\cite{Bergeard1}, where ion bombardment is used to induce two-fold in-plane magnetic anisotropy.  Those authors report no orientational melting, but the persistence of a linear domain state until fluctuations at higher temperature cause it to disappear due to the limits of time resolution in the experiment.  This suggests that, for Fe/2ML Ni/W(110), the phase transitions from the striped domain phase to either the canted phase or to the paramagnetic phase are expected to occur directly, with no intervening phase (or a very narrow one).  

Third, the dynamical transition occurs in the presence of perpendicular domains that result from the system moving to the global energy minimum.  The fact that a divergence in $\chi_{001}$ is observed shows that, as far as the system is concerned, there is a change in magnetic symmetry as the in-plane spins on the 3rd layer islands reorient to a perpendicular geometry, making the entire magnetic system perpendicular.  This means that, in a practical sense, the magnetic contribution of the spins within the domain walls is negligible.   Inverting this argument, the presence of the peak allows a calibration of the values of the ratio of domain width to domain wall width, $L/\ell$, where the system responds as if the spins in the domain walls do not break the perpendicular magnetic symmetry.

This argument can be made quantitative by using the data for $\theta_R$, $\theta_{001}$ and the magnetic constant in fig.(\ref{fig4}a) and (b), to calculate $L/\ell$ when $\theta = \theta_{001}$ using eq.(\ref{ntheta}).  The result of this calculation is shown on a logarithmic scale in fig.(\ref{fig4}c).  For comparison, the inverse of the width of the peak in $\chi_{001}$ is plotted in fig.(\ref{fig4}b), using open symbols.   At low temperature, the peak in $\chi_{001}$ is very narrow and $L/\ell$=3900 at 270 K.  Even as  $L/\ell$ falls to a value of 39 at 330 K, $\chi_{001}$ has broadened only slightly in fig.(\ref{fig4}b), indicating that a change of two orders of magnitude in the proportion of the film comprised of domain walls has not made a significant difference.  However, by 360 K, $\chi_{001}$ has broadened dramatically.  At this temperature $L/\ell\approx$10.   Somewhere in the interval $39>L/\ell>10$ the presence of the domain walls begins to break the magnetic symmetry, so that there is no longer a divergent susceptibility.

This calibration of $L/\ell$ as a function of temperature can now be applied to the measurements as a function of deposition.  For concreteness of discussion, consider any of the panels in fig.(\ref{fig2}b) through (e).  At the deposition where the dynamical transition occurs, $L/\ell$ is large and $\chi_{001}$ is divergent.  Moving to higher deposition, eq.(\ref{ntheta}) shows that $L/\ell$ gets smaller and smaller.  When $L/\ell \approx$ 10 the domain walls break the magnetic symmetry and a divergent susceptibility is no longer expected.  This occurs well before $\theta = \theta_R$, where $L/\ell \approx$ 4.3.  This illustrates quantitatively that there will be no divergence in $\chi_{001}$ at the thickness dependent re-orientation transition in the presence of the domain phase.  

These experimental data therefore support the suggestion of Pighin \textit{et al.}\cite{Pighin1}, that there is no critical phase transition at $\theta_R$ where canting begins because there is no differentiation between domain walls and domains.  This implies that there is no phase line between the perpendicular and canted domain states, although a qualitative distinction may be useful for physical arguments.  Simulations that find a phase transition line between canted domain and Ising domain phases (accompanied by, for example, a peak in the specific heat\cite{Whitehead}) might be influenced by finite size effects due to coarse graining in the simulation\cite{Whitehead2}.  Coarse graining is necessary to increase the effective size of the simulated system, but the domain walls appear Ising-like prematurely once they are thinner than the grain size.

Fourth, there is no divergence of  $\chi_{\perp}$ at the dynamical transition, even though there is a divergence in $\chi_{001}$.  This supports the argument that, because both the initial and final perpendicular domain states have $M_{\perp}$=0, this is not a useful order parameter for the transition, not withstanding questions of relative scale.  The absence of this peak at the dynamical transition implies that the absence of a divergence in $\chi_{\perp}$ at the equilibrium transition can be interpreted with confidence.  The experiments therefore support the results of  simulations of the equilibrium transition\cite{Whitehead,Pighin1}.  The simulations show that, in the neighbourhood of the in-plane state, the canted state takes the form of  a sinusoidal modulation of the magnetization with a low amplitude in the perpendicular direction.  The amplitude of the perpendicular modulation goes continuously to zero as $K_{eff}$ is reduced and the system enters the in-plane state.  Since both the sinusoidal and in-plane states have $M_{\perp}$=0, there is no critical phase transition in the magnetization.  The simulations find that the transition to the in-plane state is determined instead by the domain orientational order parameter\cite{Whitehead}, $O$, as the ordered stripe pattern disappears. 

Finally, we make some speculative comments on the transition from the perpendicular domain state to the paramagnetic state as the temperature is varied.  Consider making this thermal transition by following the path of the dashed line in fig.(\ref{fig4}a).  The variation of the fitted magnetic constant $\Gamma/\Omega b^2$ along this path due to temperature renormalization is given by the solid points in fig.(\ref{fig4}b).    For example, the literature values for bulk Fe\cite{Chikazumi} and the layer thickness of Fe/2ML Ni/W(110) films\cite{Johnston1} yield $\Gamma/(\Omega b^2)=$200, in good agreement with the fitted value for the measurements at 270 K.  The magnetic constant scales essentially as the effective exchange constant $\Gamma_{eff}(T) \sim J_{eff}(T)$, since $\Omega(M_{sat})$ is constant.   In this light,  fig.(\ref{fig4}b) appears to represent a type of ferromagnetic-to-paramagnetic transition for the perpendicular magnetization due to the reduction of $J_{eff}(T)$, with a Curie temperature $T_C$ in the neighbourhood of 305 K.  This point of view is explored by Saratz \textit{et al.}\cite{Saratz}, where they identify a ``putative" $T_C$ using magnetization curves, and by Won \textit{et al.}\cite{Won1}, who identify $T_C$ by the loss of magnetic contrast in microscopy images.

A difficulty with this interpretation is that a strong, narrow peak in $\chi_{001}$ persists above $T_C$ defined in this way.  The open points in fig.(\ref{fig4}b) show that the inverse width of $\chi_{001}$ is essentially unchanged from its value at low temperature until about 340 K.  It is not clear how the system can respond with long-range magnetic coherence in the dynamical reorientation in the temperature range of 305 to 340 K if it is in a locally disordered paramagnetic state.  It appears that the inverse peak width $\theta_{001}/\Delta \theta$ is a better qualitative indicator of the transition to microscopic paramagnetism than is $J_{eff}(T)$, and that this transition occurs near or above 360 K.

These qualitative observations offer experimental support to the computational simulations of the thermal transition from the perpendicular domain state to the paramagnetic state\cite{Whitehead,Pighin1}.  These studies find no Curie-type transition of the magnetization, but rather a continuous evolution from the mesoscopic paramagnetic behaviour of the domain state to microscopic paramagnetism.  This is again consistent with the assertion that $M_{\perp}$ is not an order parameter of the system in the critical region, but that  a second-order transition occurs in the orientational order parameter, $O(T)$, as the domain pattern symmetry changes from striped to tetragonal.  We speculate that, if the simulations were made in the presence of two-fold in-plane anisotropy as is the case in the present experiments, the second-order transition in the orientational order parameter would mark the transition from the perpendicular stripes to microscopic paramagnetism.  These are subtle questions that deserve further study.

\section{Conclusions}
As Fe/2ML Ni/W(110) films are grown they undergo three distinct versions of the same spin-reorientation transition.  At low Fe deposition, domain walls are pinned and a metastable reorientation occurs due to the local energetics of 3rd layer Fe islands.  At intermediate deposition the domain walls depin and a dynamical reorientation occurs as the system is able to access a global energetic minimum.   At even higher deposition, an equilibrium reorientation from the ordered perpendicular domain state occurs.  The dynamical reorientation is marked by a divergence in $\chi_{001}$, but no divergence in $\chi_{\perp}$, in agreement with simple models of magnetic symmetry breaking.   

A detailed expression for the equilibrium perpendicular domain width as a function of deposition (eq.(\ref{chiplot})) gives an excellent quantitative description of $\chi_{\perp}$, and has been used to determine the phase line $\theta_R(T)$ for the equilibrium transition.  This is a significant improvement on many previous studies that relied on a qualitative linear relation between the film thickness and the logarithm of the domain width.  The experimental $\theta_R(T)$ is then used to confirm the identity of the dynamical transition by fitting the peaks in $\chi_{001}$ to a quantitative model of the deposition where domain wall depinning occurs.  This identification of the dynamical transition permits a comparative analysis with the equilibrium transition.

The experimental results for the dynamical version of the reorientation clarify or corroborate the nature of the phase transitions in the equilibrium version.  First, they establish that the transition is not first-order.  Second, they demonstrate quantitatively that domain walls in the perpendicular domain state break the perpendicular magnetic symmetry once $L/\ell<10$, so that no divergence in $\chi_{001}$ is expected at the equilibrium transition between the perpendicular and canted states.  Although a distinction between the perpendicular and canted domain states is useful for physical arguments, the experiments are not consistent with an equilibrium critical phase transition between them.  Finally, they corroborate, by providing a second example, that the absence of a divergence in $\chi_{\perp}$ near $\theta_R(T)$ is a reliable null result.  This agrees with simulations showing that the equilibrium transition from the canted to in-plane state is described using the domain orientation $O$ as the order parameter,  and not by the magnetization as an order parameter.  Finally, the experimental results also offer qualitative support to the idea that the thermal transition from the domain state directly to paramagnetism is not Curie-like, but is also decribed by the domain orientation as an order parameter.

This comparative analysis provides experimental confirmation of a counter-intuitive result:  there are no divergences in the magnetic susceptibility at the equilibrium reorientation transition because neither the magnetization nor its orientation undergo a second-order critical phase transition.  For an equilibrium system studied in zero field, the dipole interactions create perpendicular domains that either remove $M_{\perp}$ as a useful order parameter, or break magnetic symmetries \textit{via} the domain walls.  Simulations predict that the second-order critical phase transitions that do occur are associated with the orientational structure of the perpendicular domain patterns themselves.  


\begin{acknowledgments}
Financial support for this work  was provided by the Natural Sciences and Engineering Research Council of Canada.
We acknowledge helpful discussions with K. De'Bell.
\end{acknowledgments}

\bibliography{re-reorientation}

\end{document}